# Comparative studies of optoelectrical properties of prominent PV materials: Halide Perovskite, CdTe, and GaAs

*Fan Zhang[1], Jose F. Castaneda[1], Shangshang Chen[2], Wuqiang Wu[2], Michael J. DiNezza[3], Maxwell Lassise[3], Wanyi Nie[4], Aditya Mohite[5], Yucheng Liu[6], Shengzhong Liu[6], Daniel Friedman[7], Henan Liu[1], Qiong Chen[1], Yong-Hang Zhang[3], Jinsong Huang[2], and Yong Zhang[1*]*

[1]Department of Electrical and Computer Engineering, The University of North Carolina at Charlotte, Charlotte, North Carolina 28223, USA

[2]Department of Applied Physical Sciences, The University of North Carolina at Chapel Hill, Chapel Hill, North Carolina 27599, USA

[3]School of Electrical, Computer and Energy Engineering, Arizona State University, Tempe, Arizona 85287, USA

[4]Materials Physics and Application Division, Los Alamos National Laboratory, Los Alamos, NM 87545, USA

[5]Department of Chemical and Biomolecular Engineering and Department of Material Science and Nanoengineering Rice University, Houston, Texas 77005, USA

[6]Key Laboratory of Applied Surface and Colloid Chemistry, National Ministry of Education; Institute for Advanced Energy Materials, School of Materials Science and Engineering, Shaanxi Normal University, Xi'an 710062, China

[7]National Renewable Energy Laboratory, Golden, Colorado 80401, USA

*Correspondence: yong.zhang@uncc.edu

Note: in this version the excitation densities were computed using measured laser profiles instead of those calculated using the diffraction limited formula.

**Abstract**

We compare three representative high performance PV materials: halide perovskite $MAPbI_3$, CdTe, and GaAs, in terms of photoluminescence (PL) efficiency, PL lineshape, carrier diffusion, and surface recombination, over multiple orders of photo-excitation density. An analytic model is used to describe the excitation density dependence of PL intensity and extract the internal PL efficiency and multiple pertinent recombination parameters. A PL imaging technique is used to obtain carrier diffusion length without using a PL quencher, thus, free of unintended influence beyond pure diffusion. Our results show that perovskite samples tend to exhibit lower Shockley-Read-Hall (SRH) recombination rate in both bulk and surface, thus higher PL efficiency than the inorganic counterparts, particularly under low excitation density, even with no or preliminary surface passivation. PL lineshape and diffusion analysis indicate that there is considerable structural disordering in the perovskite materials, and thus photo-generated carriers are not in global thermal equilibrium, which in turn suppresses the nonradiative recombination. This study suggests that relatively low point-defect density, less detrimental surface recombination, and moderate structural disordering contribute to the high PV efficiency in the perovskite. This comparative photovoltaics study provides more insights into the fundamental material science and the search for optimal device designs by learning from different technologies.

**Introduction**

The lead based organic-inorganic hybrid perovskite, such as MAPbI$_3$, has exhibited the fastest improvement in solar cell efficiency among all the known materials, since Kojima et al. first attempted to use it in photovoltaic application in 2009 [1]. It is one of the few materials that have been shown capable of achieving greater than 20% single-junction efficiency: GaAs (29.1%), Si (26.7%), InP (24.2%), CIGS (22.9%), CdTe (21.0%/22.1%), and MAPbI$_3$ (20.9%/23.7%) [2]. The material properties of MAPbI$_3$ and related structures have been studied extensively and intensively in recent years [3, 4], and various mechanisms have been proposed to explain the impressive performance of these materials in PV and other applications, in particular in their polycrystalline forms with which the best performance were typically achieved. The explanations are largely along the line of "defect tolerance", which could mean any of these: (1) defect states being either shallow relative to the band edges or in resonant with the band states [5, 6], (2) defect densities being low [7, 8], and (3) defect capture cross-sections being small [9]. Other explanations are based on the intrinsic properties of the material, for instance, low radiative recombination rate, implying a long diffusion length [10, 11], high absorption [12], band-like charge transport [13], high external photoluminescence (PL) efficiency [14-17]. Ultimately, any of these intrinsic attributions requires the defects being ineffective. We note that these considerations view the hybrid as an ordered structure, which neglects the potential impact of the disordering nature of the structure, associated with the random orientations of the organic molecules. It has been shown through electronic structure simulation that variations in the molecular orientation could lead to a large fluctuation in bandgap, from an order of 0.1 eV [18] up to 2 eV [19], simulated using small supercells. We note that the impact of the structural disordering depends on not only how it is simulated but also the material parameter of interest and method of probe [20]. The disordering-assisted "defect tolerance" is not at all unique in the perovskite. It is well known that despite high dislocation densities in the order of low $10^8$ cm$^{-2}$ in epitaxially grown In$_x$Ga$_{1-x}$N quantum well (QW) light emitting devices (LEDs), high external quantum efficiencies in electroluminescence (e.g., > 75% [21]) are readily achievable at low injection level (typically < 5 A/cm$^2$) [22]. The primary mechanism is actually the unintended structural fluctuation in the QWs that suppresses the carrier diffusion, leading to a reduced diffusion length in the order of 200 nm [23], and thus weakens non-radiative recombination loss [24]. A similar effect also exhibits in intentionally induced lateral carrier confinement in GaAs QWs [25]. This work will examine the disordering effects in the hybrid perovskites.

The hybrid perovskites were often compared against the organic semiconducting materials. Since all the other high-performance PV materials are inorganic, it would be more meaningful to compare the perovskites with those inorganic materials, which can better reveal advantages and disadvantages of the two groups, and help to improve both. After all, besides the similarity in the achieved record efficiencies, being a crystalline material, the hybrid perovskite resembles more a typical inorganic semiconductor than an organic material. Despite the existence of the large amount of literature on the perovskites, a detailed comparative study of the perovskite and conventional semiconductors under the same conditions is rare, besides cosmetic comparison of metrics. Therefore, the primary goal of this work is to offer an objective and direct comparison under the same conditions between the hybrid perovskites and two representative inorganic materials of the above-mentioned set, namely, GaAs and CdTe, in terms of a few key material properties, such as PL efficiency, PL lineshape, carrier diffusion length, and surface passivation.

PL efficiency, in particular under low excitation density close to the solar illumination, has been shown to have a positive correlation with PV efficiency [26, 27]. Internal PL quantum efficiency of 93% (external 42%) under one Sun has recently been reported for TOPO treated MAPbI$_3$ [17], approaching that of the best GaAs (internal PL quantum efficiency of 99.7%) [28]. However, even for a material like GaAs that has been studied for decades, PL efficiency can vary greatly between samples with subtle variations in their growth conditions. It is of great interest to compare perovskite samples prepared by different methods and conditions with GaAs and CdTe samples prepared under close to optimal growth conditions. Furthermore, it is important to compare the impact of the surface recombination to PL efficiency between the perovskite and inorganic materials. It is well-known that appropriate surface passivation is critical for achieving high device efficiency for the inorganic materials [29, 30]. In the perovskites, surface recombination has been shown to be significant in both carrier lifetime [31] and PL efficiency [17]. Nevertheless, the fact that high PL and device efficiencies have been achieved for the perovskites with relatively small effort seems to suggest the surface recombination in the perovskites are possibly less detrimental or easier to be passivated.

It is a common belief that long carrier diffusion length is important for superior PV performance for the perovskites or desirable in general. Although carrier diffusion lengths over 10 µm [32] even exceeding 3 mm [33] were reported for single crystalline MAPbI$_3$, the reported values for the polycrystalline samples were much shorter: ~100 nm from time-resolved PL-quenching measurements [34, 35]. These values are indeed much longer than most organic semiconductors (typically ~10 nm) [10], but much shorter than most inorganic materials with decent quality. For instance, the carrier diffusion lengths in GaAs epilayers with proper surface passivation were readily found to be ≥ 5 µm [36, 37], and could even be > 50 µm in lightly doped samples [38]. While the carrier diffusion length in a single crystalline perovskite is important as a fundamental property of the material, here we are more interested in knowing the diffusion length in a poly-crystalline perovskite sample, because the high efficiency solar cells were typically realized with the latter type and the technology is more likely to be scalable. In the literature, the carrier diffusion length was often obtained with a "quencher" of one type or another, such as a surface recombination layer or an electrode. It has been shown that the presence of a "quencher" in a semiconductor, for instance an extended defect, can alter the carrier diffusion that is supposed to be purely induced by the concentration gradient without any bias or additional driving force [37, 39]. Another significant difference between the perovskite and a conventional semiconductor like GaAs may be that the PL lineshape of the perovskite at room temperature is rather different from the latter, suggesting subtle difference in carrier thermalization within the band states [40]. A more careful comparison should be made between these materials to reveal the actual significance of the diffusion length in device performance and the implication of the differences in their spectroscopy features.

In this work, we make side-by-side comparison of PL efficiencies, over a wide range of excitation density (from around 0.01 Sun to over 4,000 Sun), for poly- and single crystalline MAPbI$_3$ samples with and without surface treatment and/or prepared by different methods, CdTe double and single heterostructures (with either top or bottom passivation layer) grown by MBE, and GaAs double heterostructures grown by MBE and MOCVD. The results are analyzed by a theoretical model to extract the internal quantum efficiency and pertinent material parameters that describe radiative and nonradiative recombination processes. A PL imaging technique with diffraction limited local excitation is used to directly probe and compare the carrier diffusion for the three materials without the complication of any intentionally introduced "quencher". Furthermore, thermal distributions of photo-generated carriers are compared between the three materials. We find that the polycrystalline perovskite can readily achieve a higher PL efficiency at the low excitation density (1 Sun or below) than CdTe and GaAs, despite that the latter two typically exhibit much larger diffusion lengths. The comparison in PL efficiency, diffusion, and PL lineshape between the three materials indicates existence of significant structural disordering and band structure distortion in the perovskite. The disordering plays an important role in suppressing nonradiative recombination, which, together with the low defect density, contributes to the high performance of perovskite solar cells.

**Materials and Methods**

MAPbI$_3$ samples from three different sources are used in this study: (1) Polycrystalline thin films of University of North Carolina at Chapel Hill group [41]: one consists of ~200 nm size grains under SEM but with smooth surface morphology under optical microscope (referred to as "UNC"), and another consists of ~250 nm size grains under SEM and 50 μm domains under optical microscope, both being about 500 nm in thickness. The second sample has an oxysalt protection layer on top for the purpose of surface passivation and protection ("UNC-passivated") [42]. (2) A polycrystalline thin film of Los Alamos National Lab group [43], about 450 nm thick and 50 μm in optical domain size ("LANL"). (3) Macroscopic size (millimeters to centimeters) single crystalline samples of Shaanxi Normal University group ("SNU") [44]. All perovskite samples were received in vacuum or with inert gas packed, and stored in a vacuum desiccator to slow the degradation process induced by humidity. For up to a few months, no observable change in PL intensity and peak position were found, which reflects significant improvement in material stability compared to the polycrystalline samples that were used in our previous studies [40].

The CdTe samples were provided by Arizona State University (ASU), all with 1 μm thick CdTe epilayer grown by MBE on InSb substrates with an InSb buffer layer. A set of three samples grown under the same conditions: one ("CdTe-DH-A1561") is a double heterostructure (DH) with 30 nm $Mg_{0.18}Cd_{0.82}Te$ top and bottom barriers, the other two have either only the top ("CdTe-Top") or bottom ("CdTe-Bottom") barrier. The first and second samples also have a 10 nm thick CdTe capping layer. These samples are used to exam the surface passivation on the inorganic semiconductors. Another CdTe DH sample ("CdTe-DH-A1671") is included with slight variations in the structure (with 15 nm $Mg_{0.5}Cd_{0.5}Te$ barriers, 30 nm CdTe cap), where a higher Mg composition is used to improve carrier confinement. Details in the material growth and device demonstration can be found elsewhere [45-47].

Two GaAs double heterojunction samples are used in this study. One is $Al_{0.4}Ga_{0.6}As/GaAs/Al_{0.4}Ga_{0.6}As$ with 1 μm GaAs and 30 nm $Al_{0.4}Ga_{0.6}As$ barriers grown by MBE by the ASU group ("GaAs-DH-B2206") [48]. The other is a $Ga_{0.51}In_{0.49}P/GaAs/Ga_{0.51}In_{0.49}P$ DH with 0.5 μm GaAs, 0.1 μm $Ga_{0.51}In_{0.49}P$ lower barrier and 5 nm upper barrier of the same grown by MOCVD by National Renewable Energy Laboratory (NREL) group ("GaAs-DH-WA540"). Details of the growth method can be found elsewhere [49].

The measurements were conducted with a Horiba LabRAM HR800 confocal Raman microscope with a 1200 g/mm grating. A 532 nm laser was used as the excitation source. Data were taken using a 4× (NA = 0.1) or 10× (NA = 0.25) microscope lens. The excitation density is estimated as D = P/A, where A is the area determined by the full width at half maximum of the measured laser profile. The spot sizes are 15.5 and 6.0 μm, respectively, for the 4× and 10× lens. The low NA lens was used to reduce the impact of carrier diffusion on the internal quantum efficiency (IQE), because the diffusion effect is equivalent to nonradiative recombination for the radiative recombination at the excitation site under the confocal mode [50]. This effect is only significant when the diffusion length is significantly larger than the laser spot size. Since the carrier diffusion length depends on the excitation density [39], the excitation spot size could potentially affect the excitation density dependence of PL. The comparison between the 4× and 10× lens confirms that with the use of the 4× lens the diffusion effect is minimal even for GaAs samples that tend to have much longer diffusion lengths than the perovskite samples (See **Figure S1** for the comparison). The laser power was measured at the exit of the microscope lens. The laser power (~16 mW of full power) was attenuated either using built-in attenuators D1–D4, approximately giving 1–4 order attenuation, or by reducing the operation current of the laser. The power density varies from below 0.01 W/cm$^2$ to around 10$^5$ W/cm$^2$. All measurements were performed at room temperature in ambient condition. For the perovskite sample, it has been widely reported in the literature [40, 51-53] that PL intensity exhibits various forms of slow time dependence, i.e., PL intensity increases gradually over a time scale up to a few hundred seconds, particularly at low excitation density. We have observed the similar effect in all perovskite samples measured. However, it typically reached a steady intensity after certain time, depending on the power density and sample (See **Figure S2** for a few typical examples). The data reported here are the stabilized intensities. For each PL spectrum or intensity shown in the main text, it is averaged over at least 8 different locations on the sample surface.

To obtain the actual laser power absorbed by the sample, surface reflection loss has been corrected by using a white light source and a UV enhanced aluminum mirror of known reflectance. The calibration was performed with a 100× (NA = 0.9) lens to collect the light of both specular and most diffuse reflection [54]. Reflectivity (R) of MAPbI$_3$, CdTe and GaAs were determined to be 0.22, 0.247 and 0.37, respectively at the excitation wavelength 532 nm, which match well with the calculated values from their refractive indexes in literatures [55-57]. All power densities mentioned are corrected values by multiplying a factor (1-R) to the laser power.

## Results and Discussion

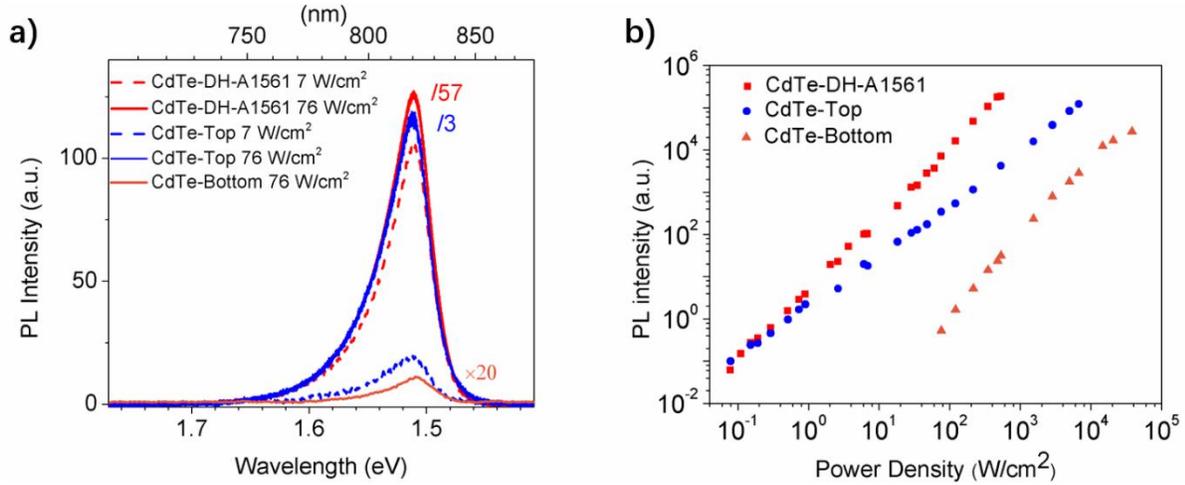

FIG.1. Excitation density dependences of three CdTe samples: CdTe-DH-A1561 (double heterostructure), CdTe-Top (with only the top barrier), and CdTe-Bottom (with only the bottom barrier). (a) Representative PL spectra of the three samples at different excitation densities. (b) Excitation density dependences of their PL intensities.

### 1. Surface effects

Firstly, we examine the influence of the surface passivation/barrier layer on the set of CdTe samples: CdTe-DH-A1561, CdTe-Top, CdTe-Bottom. The 10× lens was used for these measurements. The passivation/barrier layer serves two purposes: to passivate the surface defect states and confine the photo-generated carriers inside the active layer. **Figure 1(a)** presents several PL spectra for the CdTe samples at different excitation densities. Their PL intensities differ greatly. CdTe-DH-A1561 exhibits the highest PL intensity among all the three samples, whereas the "CdTe Top" is in between, indicating that the passivation/barrier layers are crucial and effective. For instance, at 76 W/cm$^2$, PL intensity of "CdTe Bottom" is ~1/700 and ~1/15,000 of the CdTe-Top and CdTe-DH-A1561, respectively. The results also indicate that relatively speaking, the top surface passivation is more important than the bottom surface. **Figure 1(b)** summarizes PL intensities for these CdTe samples under power densities ranging from ~0.08 to ~4×10$^4$ W/cm$^2$. Clearly, the PL signal of "CdTe Bottom" is always the weakest among the three under the same excitation density. It is interesting to note that when the power density is reduced to ~0.4 W/cm$^2$ and below, the PL intensities of the DH and top-barrier-only samples show very little difference, which suggests that at the low excitation density, the bulk Shockley–Read–Hall (SRH) dominate. However, with increasing laser density, the contrast between CdTe-DH-A1561 and CdTe-Top becomes apparent. At the excitation wavelength of 532 nm, CdTe absorption length is about 110 nm. Qualitatively, the change could be explained as that increasing excitation density leads to partial saturation of the SRH recombination loss and an increase in carrier diffusion length [39], thus, interface recombination loss at the back InSb/CdTe interface or carrier transfer into the InSb substrate becomes more significant for CdTe-Top. The comparison between the three samples suggests that the front surface recombination alone led to over three orders in magnitude reduction in PL efficiency.

In contrast, the surface recombination appears to be much less detrimental in the perovskite samples. At the 1 Sun level, the difference between the control and TOPO capped sample was found to be about a factor of 25 difference [17]. Between all our perovskite samples, with or without top surface treatment, the maximum PL intensity difference varies from around a factor of 60 at ~0.1 Sun to around 4 at ~250 Sun (to be discussed in detail later). We note that in general the surface recombination effect is relatively more important in the high illumination regime than in the low illumination where SRH is more dominant. The relatively small sample variation under high excitation suggests that surface recombination, although does exist, is less effective in the perovskite. In fact, as will be discussed next, the IQE could be as high as 80% at 50 Sun even for a non-passivated perovskite thin film, and close to 100% for a surface passivated sample.

2. **PL lineshape comparison**

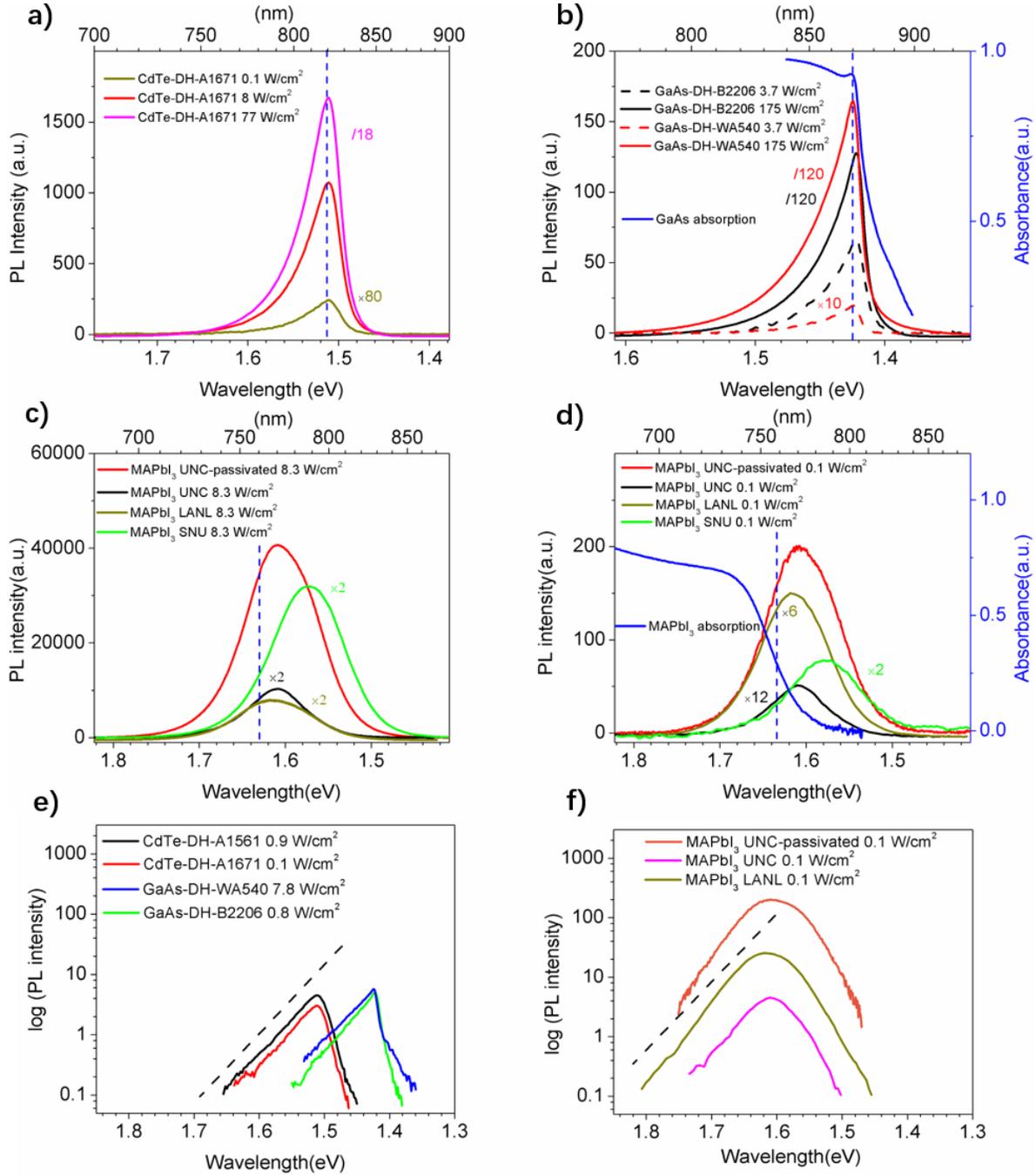

FIG.2. PL spectra of (a) CdTe-DH-A1671, (b) GaAs-DH-WA540 and B2206, (c) and (d) MAPbI$_3$ from UNC-passivated, UNC, LANL and SNU at different excitation power densities under 10× lens. Blue curve in (b) and (d) shows the absorption spectrum of GaAs and MAPbI$_3$, respectively. Vertical dashed lines indicate the positions of the excitonic bandgaps. (e) and (f) show PL spectra of all materials in semi-log scale. Dashed line is a guide for Boltzmann distribution at 300 K.

**Figure 2** compare PL spectra of the three types of samples under different excitation levels. **Figure 2(a)** plots the spectra of CdTe-DH-A1671 that exhibits considerable higher efficiency, in particular at low excitation densities, than CdTe-DH-A1561. Whether it is a double or single hetero-structure, all CdTe samples show a PL peak at around 1.511 eV that is very close to the excitonic bandgap of $E_{gx} \approx 1.510 – 1.513$ eV [58]. The full widths at half maximum (FWHM) are in the range of 44 – 45 meV. **Figure 2(b)** are the spectra of two GaAs samples: GaAs-DH-B2206 by MBE and GaAs-DH-WA540 by MOCVD. The former has a considerably higher efficiency at the low excitation region but lower at the high excitation region. The peak positions are around 1.424 eV, again very close to the excitonic bandgap at 1.425 eV [59], and the FWHM is around 30 – 35 meV. Both CdTe and GaAs PL lineshape exhibit major broadening toward the higher energy side due to thermal distribution of carriers, as expected in a conventional semiconductor [60]. Note that at room temperature despite the GaAs exciton binding energy ($E_x$) is merely 4.2 meV $\ll$ kT, excitonic absorption is visible albeit being weak, as shown by the absorption spectrum included in **Figure 2(b)** [59]. For CdTe, despite a larger $E_x$ around 10 meV, room temperature excitonic absorption is not observable because of stronger exciton-LO phonon coupling [61]. However, there is still a significant amount of excitonic contribution in room-temperature PL in CdTe [58]. Also note that the excitonic bandgaps determined by different techniques are consistent with each other typically within a few meV for a high purity conventional semiconductor. For instance, for the same piece of the GaAs sample that yielded the excitonic absorption peak at 1.425 eV shown in **Figure 2(b)**, modulation spectroscopy resulted in a bandgap of 1.422 eV [59]; for single crystalline CdTe, modulation spectroscopy resulted a bandgap of 1.513 eV [62], also very close to the excitonic bandgap determined by PL [58]. Another important indication of the PL emission being intrinsic in nature is that its peak position and lineshape do not show significant variations with excitation density except for the intensity, which is the case for CdTe and GaAs samples studied in this work.

In contrast, the PL and absorption spectra of the perovskite samples are rather different from those of the conventional semiconductors in key aspects such as:

(1) The PL peak position varies substantially from sample to sample [40], and is often significantly below the excitonic bandgap, estimated at 1.634 eV (by extrapolating from the lower temperature data) [63]. Small exciton binding energy, around 12 meV [63], is often cited as reason for not being able to observe excitonic absorption peak at room temperature in MAPbI$_3$, as shown in virtually all absorption spectra reported in the literature for the material and the one included in **Figure 2(d)** from one of our samples (UNC-passivated). However, the comparison to GaAs indicates that the small exciton binding energy does not limit the observation of excitonic absorption, but rather the disordering likely plays an important role, and phonon scattering might also contribute, which is supported by the observation that their PL linewidths are substantially larger than the inorganic counterparts, in the range of 80 – 100 meV. As a matter of fact, a small amount of impurities (e.g., 0.5% N) added to GaAs, which leads to structural disordering and electronic perturbation, can already smear or wash out the excitonic absorption peak even at 1.5 K at which the pure GaAs absorption peak is stronger than 40,000 cm$^{-1}$ [64]. In fact, absorption spectra of thin film MAPbI$_3$ could vary substantially from sample to sample (see **Figure S3** for the poly-crystalline thin-film samples studied in this work) by the standard of a well-behaved crystalline semiconductor. PL peak energies of MAPbI$_3$, as shown in **Figure 2(c)** and **(d)**, tend to be significantly lower than the estimated excitonic transition energy of 1.634 eV [63]. However, modulation (electroabsorption) or derivative spectroscopy has yielded a bandgap of 1.633 eV [65] or 1.61 eV [66] that is close to the estimated excitonic bandgap, but typically higher than one would get by using Tauc-plot, for instance, 1.598 eV from the absorption curve shown in **Figure 2(d)** [see **Figure S3(b)** for fitting]. The appearance of the derivative-like spectroscopy feature on the middle of the slope of absorption profile is a signature of a disordered semiconductor [64]. The large variation in the reported bandgaps for MAPbI$_3$ could be due to real sample variation but also deficiency in measurement. For the former, the possible strain in the small crystalline domains and/or variation in the degree of MA molecule alignment (similar to the variation in degree of ordering in a semiconductor alloy [67]) could lead to some degree of bandgap variation. For the latter, using Tauc-plot for determining bandgap on one hand will yield different bandgaps due to different spectral regions used for extrapolation [64, 66]; and on the other hand, when a thick sample is used, Tauc-plot can lead to a substantially small bandgap due to the tail absorption, which is known to the absorption measurement of a conventional semiconductor, such as GaAs [68]. The tail absorption explains the multiple reports where the PL peak energy was found to be above the "bandgap" when a thick perovskite sample was used [32, 33, 69, 70]. Clearly, Tauc-plot cannot provide a consistent way of determining the bandgap of a semiconductor.

(2) The PL lineshapes appear to be much more symmetric with respect to the peak energy, as noted earlier [40]. The broadening to the lower energy side is typically due to disordering or existence of shallow impurities or defects. An assemble of inhomogeneous but independent regions of slightly different bandgaps due to disordering often yields a Gaussian function like PL lineshape. In contrast, for a conventional semiconductor, ideally the higher energy side should reflect carrier thermalization. The $\log(I_{PL})$ vs. $1/kT$ plot of the higher energy side is customarily used for estimating the electronic temperature (or lattice temperature at low excitation density) by assuming that the thermal distribution of the carriers follows Boltzmann-like distribution $I(E) \propto (E-E_g)^{1/2}\exp[-(E-E_g)/kT]$, where E is the emission energy and $E_g$ is band gap [71]. As shown in **Figure 2(e)**, the slopes for the CdTe and GaAs samples indeed closely match that given by the 300 K slope, whereas in **Figure 2(f)** the slopes for the hybrid samples would suggest a lattice temperature below < 300 K, which is impractical. The slope reflects the electron temperature, which is often slightly higher than the lattice temperature due to unintended laser heating. It is known that perovskite has relatively poor thermal conductivity [72]. The abnormality in PL lineshape reminds us the spectroscopy signatures of a semiconductor alloy where disordering leads to the similar effects such as broadening and distortion in PL and absorption lineshape [67].

**3. PL efficiency comparison**

Since PL efficiency for a given type of material can vary greatly with changing growth conditions and surface passivation, it is not straightforward to make meaningful comparison of the PL efficiencies between different materials. By studying perovskite samples from a few groups synthesized with diverse methods [41-44], we intend to show that perovskite samples are relatively easier to achieve a high PL efficiency at low excitation density (e.g., 1 Sun), compared to the best quality CdTe and GaAs DH samples that we have identified [47, 49]. The CdTe and GaAs samples used in this study have very low densities of dislocation type extended defects (in the order of $10^3/cm^2$) from PL imaging studies [50]. Therefore, their PL efficiencies are mostly dictated by the point defects that act as SRH recombination centers and by surface passivation. To minimize the potential impact of carrier diffusion, all data for PL efficiency comparison were taken with the 4× lens. The highest excitation density used for the perovskite was around 80 W/cm$^2$, beyond that significant degradation occurred [40]. For CdTe, PL efficiency droop was observed at around 160 W/cm$^2$, possibly due to photo-induced tellurium precipitation [73]. For GaAs, PL efficiency droop may occur after reaching $10^4$ W/cm$^2$ (perhaps due to heating and other high carrier density effects), but permanent structural damage will not happen at least till $10^6$ W/cm$^2$ [39].

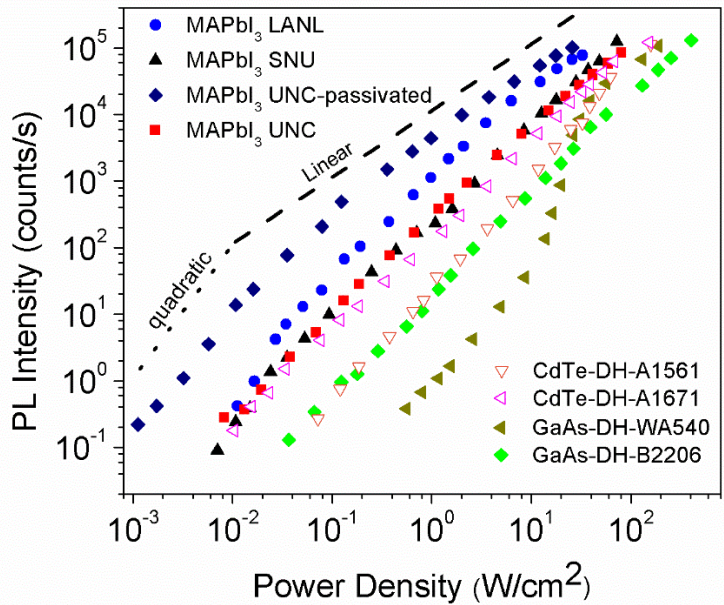

FIG.3. Excitation power density dependence of PL intensity for perovskite MAPbI$_3$ (UNC-passivated, UNC, LANL and SNU), CdTe (CdTe-DH-A1671 and CdTe-DH-A1561), and GaAs (GaAs-DH-B2206, GaAs-DH-WA540). The dashed line indicates the linear dependence slope, and dotted line for the quadratic dependence.

Here we compare the PL intensity vs. excitation density for the perovskite, CdTe, and GaAs samples described earlier under identical measurement conditions, as summarized in **Figure 3** in a log-log plot. The comparison clearly shows that the differences between the samples are most significant in the low excitation density region. Significantly, the hybrid samples almost consistently exhibit higher PL efficiencies than the inorganic ones. For instance, in the region of near 0.01 W/cm$^2$, close to the power density of 0.1 Sun, only one inorganic sample, CdTe-DH-A1671, is able to yield above-noise-level signal practically measurable by the spectroscopy system. Among the four types of perovskite samples, the polycrystalline film with oxysalt cap (UNC-passivated) shows the strongest PL signal, especially in the low power density region. The difference between two CdTe DH or two GaAs DH are also most apparent in the low excitation density region, reflecting the strong sample dependence of the SRH recombination loss, which is pertinent to the solar cell efficiency under one Sun. It is interesting to note that for GaAs, the MBE sample (B2206) is more efficient than the MOCVD sample (WA540) at the low excitation densities, likely due to less SRH recombination loss, but less efficient at high excitation density, possibly due to less effective surface passivation.

The relative PL intensities reflect the variations in PL external quantum efficiency (EQE). Ultimately, we are interested in comparing IQE rather than EQE. Referring IQE from EQE might be affected by the variation in light extraction efficiency between different type of samples [28]. This effect does exist, but contributes only to a small extent, because the variation in refractive index is relatively small: n = 2.6, 2.9, and 3.6, respectively for MAPbI$_3$, CdTe, and GaAs at their emission wavelengths. The estimated extraction efficiencies will be 3.7%, 3.0%, and 1.9%, respectively for the three types of samples (estimated with $1/(4n^2)$, without taking into account the further reduction by the NA value of the lens under confocal mode). Sample thickness may also play a role in EQE. As long as the thickness is much thicker than the absorption length, which is the case here for all the samples, one may assume that all light is absorbed. However, if the sample is too thick, vertical diffusion could lower the EQE, for instance, for the case of a macroscopic size single crystal. Therefore, on the qualitative level, the direct comparison of PL intensity is useful between most samples studied in this work.

Besides the comparison in PL intensity, the slope of the PL intensity vs. excitation density in the double-log plot can provide very useful information about the recombination mechanism [74]. Under the commonly adopted assumption that EQE is directly proportional to IQE, a linear excitation density dependence of PL intensity implies 100% IQE. This assumption could be invalid when photon-recycling is significant [28]. However, we argue that under the confocal collection mode, the photon-recycling effect is minimal, because it is known that it will take many cycles, thus a long lateral travel distance, before the photon could escape and be recycled [28]. As shown in **Figure 3**, it is quite apparent that most samples exhibit nonlinear dependence throughout the whole excitation density range, except for UNC-passivated approaching linearity at the high-density region, although to different degrees, the slopes of all samples are reduced with increasing excitation density. The results indicate that all samples measured have much lower than 100% IQE in the low excitation density region, even for the best perovskite sample. The nonlinear dependence for MAPbI$_3$ is consistent with multiple previous reports for the same material [17, 75-78]. Note that the reported over 90% IQE at 1 Sun [17] was obtained by applying the nonlinear IQE vs. EQE relationship [28] to the EQE data. As will be discussed below, the analysis of the functional dependence of the EQE vs. excitation density curve can offer more useful information than the absolute intensity itself, for instance, giving the value of IQE.

In short, from either direct comparison of the PL intensity or examining the excitation density dependence, we conclude that in general the hybrid materials are more immune from the SRH recombination loss or more defect tolerant compared to the inorganic counterparts.

**4. Diffusion length comparison**

Conceptually, a straightforward approach to investigate the diffusion of photo-generated carriers is using a tightly focused laser beam to generate carriers locally, then imaging the spatial distribution of the PL signal in the vicinity of the illumination site. This approach can be performed in either a CW [79] or a time-resolved mode [80]. The CW mode can directly yield the carrier diffusion length L, which is often more relevant to the device operation condition, whereas the time-resolved mode yields the diffusivity D that should be complemented by the measurement of the PL decay time $\tau$ to obtain the diffusion length through $L = \sqrt{(D\tau)}$. However, time resolved measurements often involve excitation densities far above the device operation condition, and the determination of the decay time to match with the diffusivity measurement could be ambiguous. Therefore, we adopt the CW mode and use a chromatically corrected 50× microscope lens with NA = 0.95 that produces a spot size of ~1.0 μm. Other details about the measurement set-up can be found in a previous publication [36].

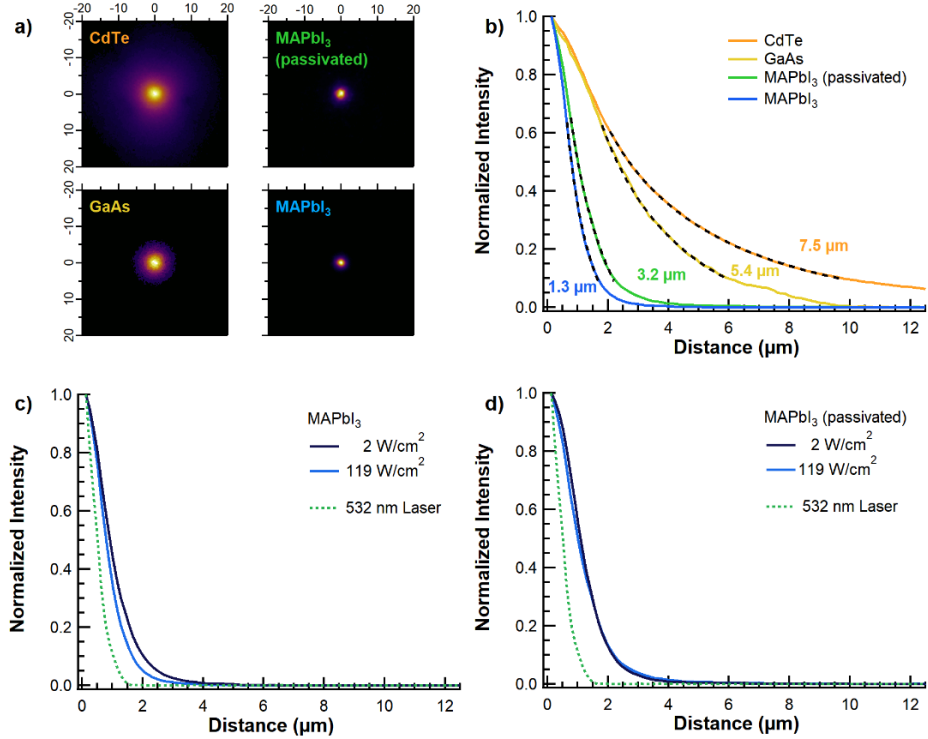

Fig. 4. (a) PL imaging results of four samples (CdTe-DH-A1617, GaAs-DH-B2206, UNC-passivated, UNC) and (b) average radial distributions, along with power density comparison for (c) UNC and (d) UNC-passivated.

Diffusion length depends on the excitation density [37, 39] and potentially even the kinetic energies of the carriers [36]. Different narrow bandpass filters can be used for probing the carriers of different kinetic energies and systematic discussion of the results of different excitation densities and imaging wavelengths will be presented elsewhere. Here we focus on the comparison of different samples under the same measurement condition. To improve detection sensitivity, a 633 nm long-pass filter is used instead of a bandpass filter to maximize collection of all emitted photons. **Figure 4** shows PL imaging results of four samples: UNC-passivated, UNC, CdTe-DH-A1671, and GaAs-DH-B2206 at 119 W/cm$^2$. As shown evidently in **Figure 4(a)**, the emitted photons in the perovskite samples are mostly from the region near the excitation site while the two inorganic samples show light emission from carriers that have diffused further away from the illumination site. The spatial profile of the PL intensity away from the illumination site can be described by the modified Bessel function $K_0(\rho/L_\tau)$ under a 2D approximation [36], where $\rho$ is the radius from the illumination site, and $L_\tau = \sqrt{(D\tau)}$ is the diffusion length defined in the diffusion equation with $1/\tau$ being the total recombination rate. The diffusion length $L_\tau$ can be extracted from fitting the angularly averaged radial distribution, as demonstrated in **Figure 4(b)**, and obtained $L_\tau$ values are 1.3, 3.2, 5.4, and 7.5 µm for UNC, UNC-passivated, GaAs-DH-B2206, and CdTe-DH-A1671, respectively, all at 119 W/cm$^2$. Note that the radial decay of the PL intensity described by $K_0(\rho/L_\tau)$ tends to be faster than the commonly adopted simple exponential decay function $\exp(-\rho/L_{exp})$, which will result in a shorter apparent diffusion length if the profile is fit to the exponential decay. The 1/e points yield $L_{exp} = 0.7$, 1.1, 2.5, and 3.4 µm, respectively, for the same data of **Figure 4(b)**. To examine the potential excitation density dependence, we compare for the perovskite samples between 119 W/cm$^2$ and 2 W/cm$^2$, and find that the variations are relatively small, as shown in **Figure 4(c)** and **(d)**. We can conclude that the diffusion lengths of the polycrystalline perovskite samples are much shorter compared to the inorganic samples, which, however, does not prevent the perovskite samples similar to those studied here to achieve high PV efficiencies around 20%. For the inorganic samples, despite exhibiting much larger diffusion lengths than the perovskite samples for instance at 119 W/cm$^2$, their PL intensities are much weaker. Thus, for the inorganic samples, the diffusion lengths are still limited by the SRH defects even in such a moderately high excitation. In contrast, for the perovskite samples, the diffusion lengths are instead largely limited by the disordering due to both microscopic domain structures [81] and intrinsic molecular random orientations. The difference between the two perovskite samples could be due to surface passivation.

Because the absorber layer thickness in a typical perovskite solar cell is around 0.5 µm, a 1-2 µm diffusion length is more than adequate for carrier collection along the vertical direction, and longer diffusion lengths would in fact increase the lateral diffusion loss through SRH recombination. In this context, the carrier diffusion lengths for the direct bandgap inorganic materials like CdTe and GaAs are probably longer than what are preferred for the optimal PV performance. If the diffusion length could be moderately reduced but without introducing additional defects, the PV performance of the inorganic materials might even be improved.

The PL imaging method implicitly assumes that photon-recycling does not contribute appreciably to the emission away from the excitation site, which is valid in most realistic situations. Photon-recycling effect can be significant only in the case of close to unity IQE. Even in that case, it typically takes a few tens of cycles, corresponding to a distance greater than the carrier diffusion length, for a re-emitted photon to escape from the front surface. Our data for the perovskite samples apparently do not yield much emission beyond the excitation site, which indicates both short diffusion length and negligible photon-recycling.

## 5. Discussion

To analyze the power dependent PL intensity in **Figure** 3 more quantitatively, we apply a three-level model that can adequately capture the physical processes of SRH centers under different generation levels [82]. The model includes two levels representing the band edge states of the conduction and valence band, respectively, and the other mimicking the "defect" states that are below the conduction band and can capture the carriers from the conduction band, then through which recombine with holes in the valence band. We can write two rate equations as:

$$\frac{dn}{dt} = G - nW_r - n\gamma_t N_t(1-f) + e_t f N_t$$

$$\frac{dN}{dt} = n\gamma_t N_t(1-f) - f N_t(W_t + e_t) \quad (1)$$

where n is the conduction band electron density, N the defect state electron density, $N_t$ the total defect state density, $f = N/N_t$ the fraction of the occupied defect states, G the generation rate, which is proportional to laser power density P, $W_r$ the radiative recombination rate, $\gamma_t$ the defect capture coefficient with $c_t = \gamma_t N_t$ the maximum capture rate, $e_t$ the re-emission rate from the defect states to conduction band, $W_t$ the defect recombination rate. Similar rate equation approaches have been used to analyze the CW PL excitation density dependence [58, 82], including some on the perovskite [75, 76, 78]. Here we adopt the approach of Ref. [82] that allows us to extract the internal quantum efficiency (IQE) $\eta$ from the excitation density dependence of CW PL.

Steady state solutions of n and N leads to the following formula for $\eta$ and $f$:

$$\eta = \frac{nW_r}{G} = \frac{1}{2}\left(1 - \frac{\alpha+\beta}{G} + \sqrt{(1+\frac{\alpha+\beta}{G})^2 - \frac{4\beta}{G}}\right), \quad (2)$$

$$f = \frac{N}{N_t} = \frac{G}{2\beta}\left(1 + \frac{\alpha+\beta}{G} - \sqrt{(1+\frac{\alpha+\beta}{G})^2 - \frac{4\beta}{G}}\right), \quad (3)$$

where $\alpha = W_r(W_t+e_t)/\gamma_t$ and $\beta = N_t W_t$. $\beta$ represents the maximum recombination rate of the defect states, and $\alpha$ describes coupling and competition between the radiative recombination and non-radiative recombination of the defect states through the capture process. This result implies in a general case a nonlinear relationship between PL efficiency and the excitation density, and it is inappropriate to write the total recombination rate as a sum of the radiative and non-radiative recombination rates, unless $f \ll 1$. To describe the relative EQE, $\eta_{EQE} = I_{PL}/P$, where $I_{PL}$ is PL intensity (in count/s), and P is the excitation power density (in W/cm$^2$), we rewrite Eq. (2) as

$$\eta_{EQE} = \frac{C}{2}\left(1 - \frac{\alpha'+\beta'}{P} + \sqrt{(1+\frac{\alpha'+\beta'}{P})^2 - \frac{4\beta'}{P}}\right) \quad (4)$$

where G is replaced by $\xi P$, $\xi$ is a constant, $\alpha$ and $\beta$ by $\alpha' = \alpha/\xi$ and $\beta' = \beta/\xi$, and 1/2 by C/2 with C being a scaling constant that depends on the PL collection efficiency. By fitting experimental data of $I_{PL}/P$ vs. P, one can obtain IQE vs. P through IQE = $\eta_{EQE}/C$. For instance, if IQE = 100% (when $\beta = 0$), those terms inside parentheses (…) add to 2. Alternatively, as P is sufficiently large such that the defect states are saturated, (…) also approaches 2. Thus, Eq. (4) can be used to obtain the IQE curve as a function of excitation density, without having to explicitly determine the carrier density or directly measure the EQE then convert it to IQE. In general, α and β can be P dependent through $W_r$ and $W_t$. If the recombination is bimolecular, we have $W_r$ ($W_t$) $\propto p = p_0 + \delta p$ for a p-type material. It is impractical to obtain an analytic solution for $\delta p$, however, it is expected to be P dependent. Therefore, we write $W_r = W_{r0}(1+dP^\zeta)$ and $W_t = W_{t0}(1+dP^\zeta)$, where $W_{r0} = Bp_0$ with B being radiative recombination coefficient, and $\delta p/p_0 = dP^\zeta$ with d and $\zeta$ being treated as fitting parameters. Note that in the simplest case of bimolecular recombination without trapping states, $\zeta = 0.5$. As a result, we can re-write Eq. (4) as:

$$\eta_{EQE} = \frac{C}{2}\left(1 - \beta_0'(1+dx^\zeta)\frac{u(1+dP^\zeta)+1}{P} + \sqrt{(1+\beta_0'(1+dP^\zeta)\frac{u(1+dP^\zeta)+1}{P})^2 - \frac{4\beta_0'(1+dP^\zeta)}{P}}\right) \quad (5)$$

where u is defined to be $\alpha_0/\beta_0 \approx W_{r0}/c_t$, assuming $W_t \gg e_t$. Note that when writing Eq. (1), we implicitly assume the material is p-type. However, the results are equally applicable to n-type, simply by replacing n with p, and interpreting the parameters accordingly.

The PL intensity data in **Figure** 3 are fitted with Eq. (5). After obtaining scaling parameter C, the data of **Figure 3** are re-plotted to yield IQE curves for those samples, as shown in **Figure 5** together with the fitted curves. The key fitting parameters are summarized in **Table 1**. Despite very different excitation density dependences between the samples, the model is able to fit most samples rather well over an excitation density range of about four orders in magnitude, for instance, from below 0.01 Sun to over 250 Sun for the perovskite sample UNC-passivated. In **Figure 5(a)**, the more efficient CdTe sample, DH-A1671, exhibits efficiency of ~2.9% IQE at 0.1 W/cm$^2$ and ~50% at 60 W/cm$^2$. However, increasing the density further to ~160W/cm$^2$ (the last data point in **Figure 5(a)**, the efficiency starts to reduce, likely due to material degradation [73]. In **Figure 5(b)**, GaAs-DH-B2206 only has ~1.2% IQE at 0.1 W/cm$^2$ but reaches ~70% at ~400 W/cm$^2$. Different from the inorganic samples, in **Figure 5(c)**, the thin film perovskite samples tend to show higher efficiencies in the low-density region and sharper increase in efficiency below 30 W/cm$^2$. The most efficient one, UNC-passivated, shows ~70% IQE at 0.1 W/cm$^2$ and already reaches the saturation point (i.e., 100% IQE) at around 2 W/cm$^2$. For the non-passivated UNC sample, it reaches ~80% IQE at 50 W/cm$^2$. The relatively lower efficiency of the single crystal perovskite (SNU) and its more complex excitation density dependence may be related to carrier diffusion into the bulk or more surface defects [31]. Nevertheless, it still reaches ~40% IQE at 70 W/cm$^2$, the same level as the better CdTe sample.

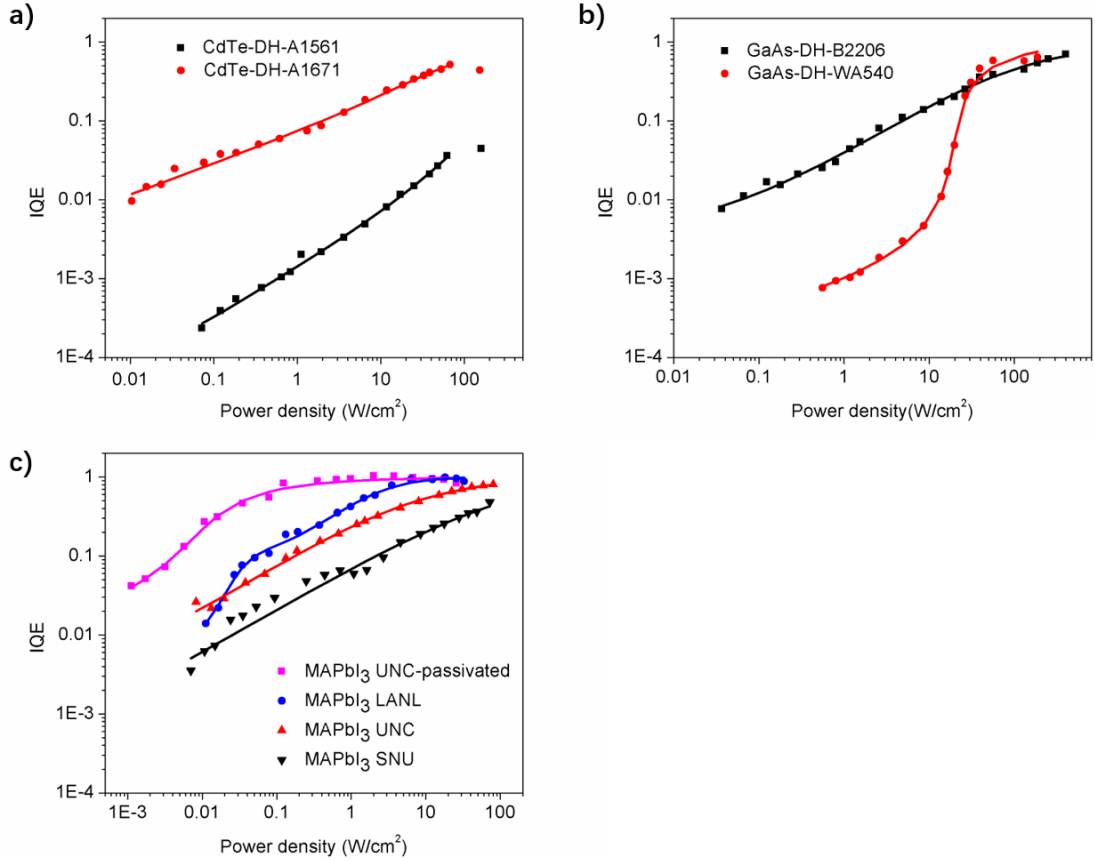

FIG.5. Excitation density dependent internal quantum efficiency for the same data of Fig. 3. (a) CdTe samples, (b) GaAs samples, and (c) MAPbI$_3$ samples. Solid lines are fitted curves.

Table 1: Fitting parameters to Eq. (5).

| Parameters | CdTe-DH-A1561 | CdTe-DH-A1671 | GaAs-DH-WA540 | GaAs-DH-B2206 | MAPbI$_3$ UNC-passivated | MAPbI$_3$ UNC | MAPbI$_3$ LANL | MAPbI$_3$ SNU |
|---|---|---|---|---|---|---|---|---|
| u | 2.4E-5 | 1.6E-4 | 6.6E-5 | 4.0E-3 | 1.1E-2 | 2.2E-3 | 3.6E-3 | 5.2E-5 |
| $\beta'_0$(W/cm$^2$) | 0.16 | 0.02 | 0.50 | 1.37 | 1.5E-3 | 5.7 | 1.2E-2 | 15.3 |
| d | 53.3 | 443.5 | 12.2 | 8.6 | 79.6 | 138.5 | 145.7 | 1410.7 |
| ζ | 0.63 | 0.40 | 0.38 | 0.63 | 0.55 | 0.59 | 1.25 | 0.54 |

More insights can be achieved from quantitative comparison of the fitting parameters between the samples. As revealed by Eq. (2), $\beta$ (= $W_t N_t$) ≠ 0 is the direct cause for IQE < 1, and $\alpha \gg \beta$ is the condition for high IQE. A large $\alpha \approx (W_r/c_t)\beta$ requires a large radiative recombination rate $W_r$ or a small capture rate $c_t$. Therefore, in a doped material where PL is primarily minority carrier recombination or the d term in Eq. (5) is small, a large ratio $u = \alpha_0/\beta_0 \approx W_{r0}/c_t$ is important for having high IQE at low excitation density and fast building up with increasing excitation density. To be specific, as it can be seen from **Table 1**, u value of CdTe-DH-A1671 is one order of magnitude larger than CdTe-DH-A1561, suggesting that the traps are less effective in the former. However, the difference could be either in the background doping level (in $W_{r0}$) or the defect density (in $c_t$). At the same time, u value of GaAs-DH-WA540 is much smaller than that of GaAs-DH-B2206, which results in the lower PL efficiency of the former in the low power density region (1-20 W/cm$^2$). The comparison suggests that the defects in the MBE and MOCVD samples could be rather different. When comparing between the inorganic samples with the hybrid perovskite samples, in most cases, u and $\beta_0$ values of the inorganic ones are smaller and larger, respectively, than those of the perovskites, with only one exception of the single crystalline perovskite sample. Thus, we have reached an important conclusion that the inorganic material, even with good surface passivation, tends to have a much higher defect capture rate in the bulk, relative to the radiative emission rate, than the poly-crystalline perovskite sample even without surface passivation. If further assuming the parameter B and $\gamma_t$ do not vary strongly between these materials, we could claim that the most important differences between them are the effective SRH type defect density, which tends to be much lower in the high quality hybrid perovskites. We suggest that the lower carrier mobility, manifested as the smaller carrier diffusion length, plays an important role in suppressing the defect capture rate in the poly-crystalline perovskite sample, thus, contributes to the high PL, possibly also PV efficiency.

Above analyses were carried out under the assumption that EQE is linearly proportional to IQE, as commonly assumed. However, if the photon recycling effect is significant, a nonlinear relationship is expected [28]. In fact, the IQE curve for the most efficient perovskite sample (UNC-passivated) shown in **Figure 5(c)** is similar to the normalized EQE curve reported for the TOPO passivated MAPbI3 sample with an IQE of 92% under 1 Sun equivalent (60 mW/cm$^2$ for 532 nm). However, there the normalized EQE under 1 Sun equivalent was ~35% [17]. In our case, fitting the data with Eq. (5) yields an IQE of ~60% under 1 Sun equivalent. Although, as pointed out above, under confocal collection mode, the photon-recycling effect should be minimal, we can nevertheless perform the similar fitting to examine what would happen if the photo-recycling effect should exist. Then the IQE would go up from 60% to 85% under 1 Sun equivalent. Details are given in the supplemental materials and shown in **Figure S4**. This exercise indicates that the UNC-passivated represents the perovskite samples of the highest PL efficiency.

The fact that a high PL efficiency at the low excitation density is relatively easy to achieve explains why with relatively simply device structures and limited optimization effort a perovskite solar cell can regularly reach above 20% efficiencies, whereas a typically lower PL efficiency at the low excitation density explains why with relatively complex device structures and long term optimization effort in material growth and device design it is still not so easy to achieve above 20% efficiency for the inorganic cells.

**Conclusions**

We present an objective, side-by-side, comparison on the optical properties of three representative high performance PV materials, including hybrid perovskite MAPbI3, CdTe, and GaAs. It is relatively easy for the poly-crystalline perovskite samples to exhibit higher PL efficiency than the inorganic counterparts, in particular under low excitation density, despite the inorganic materials tend to have longer carrier diffusion lengths. Relatively speaking, the perovskites are more immune to the surface recombination than the inorganic materials, and exhibit less effective SRH type non-radiative recombination, which together helps to achieve high PV efficiency. Comparison in PL spectrum, including relative PL peak position to the excitonic transition energy, linewidth, and spectral shape, between the three materials suggests the existence of significant intrinsic (i.e., random orientation of the molecules) and extrinsic (micro-grains) disordering effect in the hybrid perovskite. We point out that moderate disordering is actually beneficial for low non-radiative carrier recombination in achieving high luminescence and solar cell efficiency. This study provides new insights towards comprehensive understanding of halide perovskite materials and improving device performance for both hybrid and inorganic materials.

**Acknowledgement**

The work at UNC-Charlotte and UNC-Chapel Hill was supported by University of North Carolina's Research Opportunities Initiative (UNC ROI) through Center of Hybrid Materials Enabled Electronic Technology and ARO/Electronics (Grant No. W911NF-16-1-0263). The work at ASU was partially supported by AFOSR (Grant No. FA9550-12-1-0444 and FA9550-15-1-0196), Science Foundation Arizona (Grant No. SRG 0339-08), and NSF (Grant No. 1002114), the Department of Energy through Bay Area Photovoltaic Consortium (Grant No. DE-EE0004946) and Energy Efficiency and Renewable Energy (Grant No. DE-EE0007552). M.J.D. was supported by the National Science Foundation Graduate Research Fellowship (Grant No. DGE-0802261). The work at NREL was authorized in part by Alliance for Sustainable Energy, LLC, the manager and operator of NREL for the U.S. Department of Energy (DOE) under Contract No. DE-AC36-08GO28308. Funding provided by U.S. Department of Energy Office of Energy Efficiency and Renewable Energy Solar Energy Technologies Office.

**Contributions**
UNCC group (F.Z., JF. C, Y.Z) designed the project and performed the optical experiments. They wrote the manuscript, with inputs and comments from all other authors. UNC-CH group (S.C., W.W. and J.H.), LANL group (W.N. and A.M.), and SNU group (Y.L. and S.L.) synthesized MAPbI3 samples; ASU group (MJ.D., M.L., YH.Z.) grew CdTe and MBE GaAs samples; NREL group (D.F.) grew MOCVD GaAs samples.

**Conflict of Interest**
The authors declare no conflict of interest.

# References

[1] A. Kojima, K. Teshima, Y. Shirai, T. Miyasaka, Organometal Halide Perovskites as Visible-Light Sensitizers for Photovoltaic Cells, Journal of the American Chemical Society, 131 (2009) 6050-6051.

[2] M.A. Green, Y. Hishikawa, E.D. Dunlop, D.H. Levi, J. Hohl-Ebinger, M. Yoshita, A.W.Y. Ho-Baillie, Solar cell efficiency tables (Version 53), Progress in Photovoltaics: Research and Applications, 27 (2019) 3-12.

[3] J.-P. Correa-Baena, M. Saliba, T. Buonassisi, M. Grätzel, A. Abate, W. Tress, A. Hagfeldt, Promises and challenges of perovskite solar cells, Science, 358 (2017) 739.

[4] D.A. Egger, A. Bera, D. Cahen, G. Hodes, T. Kirchartz, L. Kronik, R. Lovrincic, A.M. Rappe, D.R. Reichman, O. Yaffe, What Remains Unexplained about the Properties of Halide Perovskites?, Advanced Materials, 30 (2018) 1800691.

[5] W.-J. Yin, T. Shi, Y. Yan, Unusual defect physics in CH3NH3PbI3 perovskite solar cell absorber, Applied Physics Letters, 104 (2014) 063903.

[6] H. Shi, M.-H. Du, Shallow halogen vacancies in halide optoelectronic materials, Physical Review B, 90 (2014) 174103.

[7] G. Xing, N. Mathews, S.S. Lim, N. Yantara, X. Liu, D. Sabba, M. Grätzel, S. Mhaisalkar, T.C. Sum, Low-temperature solution-processed wavelength-tunable perovskites for lasing, Nature Materials, 13 (2014) 476.

[8] V. D'Innocenzo, G. Grancini, M.J.P. Alcocer, A.R.S. Kandada, S.D. Stranks, M.M. Lee, G. Lanzani, H.J. Snaith, A. Petrozza, Excitons versus free charges in organo-lead tri-halide perovskites, Nat Commun, 5 (2014) 3586.

[9] R.E. Brandt, J.R. Poindexter, P. Gorai, R.C. Kurchin, R.L.Z. Hoye, L. Nienhaus, M.W.B. Wilson, J.A. Polizzotti, R. Sereika, R. Žaltauskas, L.C. Lee, J.L. MacManus-Driscoll, M. Bawendi, V. Stevanović, T. Buonassisi, Searching for "Defect-Tolerant" Photovoltaic Materials: Combined Theoretical and Experimental Screening, Chemistry of Materials, 29 (2017) 4667-4674.

[10] T.C. Sum, N. Mathews, Advancements in perovskite solar cells: photophysics behind the photovoltaics, Energy & Environmental Science, 7 (2014) 2518-2534.

[11] P. Azarhoosh, S. McKechnie, J.M. Frost, A. Walsh, M.v. Schilfgaarde, Research Update: Relativistic origin of slow electron-hole recombination in hybrid halide perovskite solar cells, APL Materials, 4 (2016) 091501.

[12] S. De Wolf, J. Holovsky, S.-J. Moon, P. Löper, B. Niesen, M. Ledinsky, F.-J. Haug, J.-H. Yum, C. Ballif, Organometallic Halide Perovskites: Sharp Optical Absorption Edge and Its Relation to Photovoltaic Performance, The Journal of Physical Chemistry Letters, 5 (2014) 1035-1039.

[13] L.M. Herz, Charge-Carrier Mobilities in Metal Halide Perovskites: Fundamental Mechanisms and Limits, ACS Energy Letters, 2 (2017) 1539-1548.

[14] F. Deschler, M. Price, S. Pathak, L.E. Klintberg, D.-D. Jarausch, R. Higler, S. Hüttner, T. Leijtens, S.D. Stranks, H.J. Snaith, M. Atatüre, R.T. Phillips, R.H. Friend, High Photoluminescence Efficiency and Optically Pumped Lasing in Solution-Processed Mixed Halide Perovskite Semiconductors, The Journal of Physical Chemistry Letters, 5 (2014) 1421-1426.

[15] J.M. Richter, M. Abdi-Jalebi, A. Sadhanala, M. Tabachnyk, J.P.H. Rivett, L.M. Pazos-Outón, K.C. Gödel, M. Price, F. Deschler, R.H. Friend, Enhancing photoluminescence yields in lead halide perovskites by photon recycling and light out-coupling, Nature Communications, 7 (2016) 13941.
[16] W. Tress, Perovskite Solar Cells on the Way to Their Radiative Efficiency Limit – Insights Into a Success Story of High Open-Circuit Voltage and Low Recombination, Advanced Energy Materials, 7 (2017) 1602358.
[17] I.L. Braly, D.W. deQuilettes, L.M. Pazos-Outón, S. Burke, M.E. Ziffer, D.S. Ginger, H.W. Hillhouse, Hybrid perovskite films approaching the radiative limit with over 90% photoluminescence quantum efficiency, Nature Photonics, 12 (2018) 355-361.
[18] A.M.A. Leguy, P. Azarhoosh, M.I. Alonso, M. Campoy-Quiles, O.J. Weber, J. Yao, D. Bryant, M.T. Weller, J. Nelson, A. Walsh, M. van Schilfgaarde, P.R.F. Barnes, Experimental and theoretical optical properties of methylammonium lead halide perovskites, Nanoscale, 8 (2016) 6317-6327.
[19] M. Wierzbowska, J.J. Meléndez, D. Varsano, Breathing bands due to molecular order in $CH_3NH_3PbI_3$, Computational Materials Science, 142 (2018) 361-371.
[20] Y. Zhang, L.-W. Wang, Global electronic structure of semiconductor alloys through direct large-scale computations for III-V alloys $Ga_xIn_{1-x}P$, Physical Review B, 83 (2011) 165208.
[21] Y. Narukawa, M. Sano, T. Sakamoto, T. Yamada, T. Mukai, Successful fabrication of white light emitting diodes by using extremely high external quantum efficiency blue chips, physica status solidi (a), 205 (2008) 1081-1085.
[22] J. Cho, E.F. Schubert, J.K. Kim, Efficiency droop in light-emitting diodes: Challenges and countermeasures, Laser & Photonics Reviews, 7 (2013) 408-421.
[23] D. Cherns, S.J. Henley, F.A. Ponce, Edge and screw dislocations as nonradiative centers in InGaN/GaN quantum well luminescence, Applied Physics Letters, 78 (2001) 2691-2693.
[24] S.F. Chichibu, A.C. Abare, M.P. Mack, M.S. Minsky, T. Deguchi, D. Cohen, P. Kozodoy, S.B. Fleischer, S. Keller, J.S. Speck, J.E. Bowers, E. Hu, U.K. Mishra, L.A. Coldren, S.P. DenBaars, K. Wada, T. Sota, S. Nakamura, Optical properties of InGaN quantum wells, Materials Science and Engineering: B, 59 (1999) 298-306.
[25] Y. Zhang, M.D. Sturge, K. Kash, B.P. van der Gaag, A.S. Gozdz, L.T. Florez, J.P. Harbison, Temperature dependence of luminescence efficiency, exciton transfer, and exciton localization in $GaAs/Al_xGa_{1-x}As$ quantum wires and quantum dots, Physical Review B, 51 (1995) 13303-13314.
[26] E. Yablonovitch, O.D. Miller, S. Kurtz, The opto-electronic physics that broke the efficiency limit in solar cells, in: 2012 38th IEEE photovoltaic specialists conference, IEEE, 2012, pp. 001556-001559.
[27] M.A. Green, Radiative efficiency of state-of-the-art photovoltaic cells, Progress in Photovoltaics: Research and Applications, 20 (2012) 472-476.
[28] I. Schnitzer, E. Yablonovitch, C. Caneau, T.J. Gmitter, Ultrahigh spontaneous emission quantum efficiency, 99.7% internally and 72% externally, from AlGaAs/GaAs/AlGaAs double heterostructures, Applied Physics Letters, 62 (1993) 131-133.
[29] A.G. Aberle, Surface passivation of crystalline silicon solar cells: a review, Progress in Photovoltaics: Research and Applications, 8 (2000) 473-487.
[30] K.A. Bertness, S.R. Kurtz, D.J. Friedman, A.E. Kibbler, C. Kramer, J.M. Olson, 29.5%‐efficient GaInP/GaAs tandem solar cells, Applied Physics Letters, 65 (1994) 989-991.

[31] Y. Yang, M. Yang, David T. Moore, Y. Yan, Elisa M. Miller, K. Zhu, Matthew C. Beard, Top and bottom surfaces limit carrier lifetime in lead iodide perovskite films, Nature Energy, 2 (2017) 16207.

[32] D. Shi, V. Adinolfi, R. Comin, M. Yuan, E. Alarousu, A. Buin, Y. Chen, S. Hoogland, A. Rothenberger, K. Katsiev, Y. Losovyj, X. Zhang, P.A. Dowben, O.F. Mohammed, E.H. Sargent, O.M. Bakr, Low trap-state density and long carrier diffusion in organolead trihalide perovskite single crystals, Science, 347 (2015) 519-522.

[33] Q. Dong, Y. Fang, Y. Shao, P. Mulligan, J. Qiu, L. Cao, J. Huang, Electron-hole diffusion lengths > 175 μm in solution-grown $CH_3NH_3PbI_3$ single crystals, Science, 347 (2015) 967.

[34] S.D. Stranks, G.E. Eperon, G. Grancini, C. Menelaou, M.J.P. Alcocer, T. Leijtens, L.M. Herz, A. Petrozza, H.J. Snaith, Electron-Hole Diffusion Lengths Exceeding 1 Micrometer in an Organometal Trihalide Perovskite Absorber, Science, 342 (2013) 341-344.

[35] G. Xing, N. Mathews, S. Sun, S.S. Lim, Y.M. Lam, M. Grätzel, S. Mhaisalkar, T.C. Sum, Long-Range Balanced Electron- and Hole-Transport Lengths in Organic-Inorganic $CH_3NH_3PbI_3$, Science, 342 (2013) 344-347.

[36] S. Zhang, L.Q. Su, J. Kon, T. Gfroerer, M.W. Wanlass, Y. Zhang, Kinetic energy dependence of carrier diffusion in a GaAs epilayer studied by wavelength selective PL imaging, Journal of Luminescence, 185 (2017) 200-204.

[37] F. Chen, Y. Zhang, T.H. Gfroerer, A.N. Finger, M.W. Wanlass, Spatial resolution versus data acquisition efficiency in mapping an inhomogeneous system with species diffusion, Scientific reports, 5 (2015) 10542.

[38] A. Nemcsics, K. Somogyi, Correlation between diffusion length and Hall mobility in different GaAs epitaxial layers, in:  ASDAM 2000. Conference Proceedings. Third International EuroConference on Advanced Semiconductor Devices and Microsystems (Cat. No.00EX386), 2000, pp. 265-268.

[39] T.H. Gfroerer, Y. Zhang, M.W. Wanlass, An extended defect as a sensor for free carrier diffusion in a semiconductor, Applied Physics Letters, 102 (2013) 012114.

[40] Q. Chen, H. Liu, H.-S. Kim, Y. Liu, M. Yang, N. Yue, G. Ren, K. Zhu, S. Liu, N.-G. Park, Y. Zhang, Multiple-Stage Structure Transformation of Organic-Inorganic Hybrid Perovskite $CH_3NH_3PbI_3$, Physical Review X, 6 (2016) 031042.

[41] W.-Q. Wu, Z. Yang, P.N. Rudd, Y. Shao, X. Dai, H. Wei, J. Zhao, Y. Fang, Q. Wang, Y. Liu, Y. Deng, X. Xiao, Y. Feng, J. Huang, Bilateral alkylamine for suppressing charge recombination and improving stability in blade-coated perovskite solar cells, Science Advances, 5 (2019) eaav8925.

[42] S. Yang, S. Chen, E. Mosconi, Y. Fang, X. Xiao, C. Wang, Y. Zhou, Z. Yu, J. Zhao, Y. Gao, F. De Angelis, J. Huang, Stabilizing halide perovskite surfaces for solar cell operation with wide-bandgap lead oxysalts, Science, in press (2019).

[43] W. Nie, H. Tsai, R. Asadpour, J.-C. Blancon, A.J. Neukirch, G. Gupta, J.J. Crochet, M. Chhowalla, S. Tretiak, M.A. Alam, H.-L. Wang, A.D. Mohite, High-efficiency solution-processed perovskite solar cells with millimeter-scale grains, Science, 347 (2015) 522-525.

[44] Y. Liu, Z. Yang, D. Cui, X. Ren, J. Sun, X. Liu, J. Zhang, Q. Wei, H. Fan, F. Yu, X. Zhang, C. Zhao, S. Liu, Two-Inch-Sized Perovskite $CH_3NH_3PbX_3$ (X = Cl, Br, I) Crystals: Growth and Characterization, Advanced Materials, 27 (2015) 5176-5183.

[45] S. Liu, X.-H. Zhao, C.M. Campbell, M.B. Lassise, Y. Zhao, Y.-H. Zhang, Carrier lifetimes and interface recombination velocities in CdTe/MgxCd1−xTe double heterostructures with different Mg compositions grown by molecular beam epitaxy, Applied Physics Letters, 107 (2015) 041120.
[46] Y. Zhao, M. Boccard, S. Liu, J. Becker, X.-H. Zhao, C.M. Campbell, E. Suarez, M.B. Lassise, Z. Holman, Y.-H. Zhang, Monocrystalline CdTe solar cells with open-circuit voltage over 1 V and efficiency of 17%, Nature Energy, 1 (2016) 16067.
[47] M.J. DiNezza, X.-H. Zhao, S. Liu, A.P. Kirk, Y.-H. Zhang, Growth, steady-state, and time-resolved photoluminescence study of CdTe/MgCdTe double heterostructures on InSb substrates using molecular beam epitaxy, Applied Physics Letters, 103 (2013) 193901.
[48] J.-B. Wang, D. Ding, S.R. Johnson, S.-Q. Yu, Y.-H. Zhang, Determination and improvement of spontaneous emission quantum efficiency in GaAs/AlGaAs heterostructures grown by molecular beam epitaxy, physica status solidi (b), 244 (2007) 2740-2751.
[49] M.A. Steiner, J.F. Geisz, I. García, D.J. Friedman, A. Duda, S.R. Kurtz, Optical enhancement of the open-circuit voltage in high quality GaAs solar cells, Journal of Applied Physics, 113 (2013) 123109.
[50] C. Hu, Q. Chen, F. Chen, T.H. Gfroerer, M.W. Wanlass, Y. Zhang, Overcoming diffusion-related limitations in semiconductor defect imaging with phonon-plasmon-coupled mode Raman scattering, Light: Science & Applications, 7 (2018) 23.
[51] Y. Tian, M. Peter, E. Unger, M. Abdellah, K. Zheng, T. Pullerits, A. Yartsev, V. Sundström, I.G. Scheblykin, Mechanistic insights into perovskite photoluminescence enhancement: light curing with oxygen can boost yield thousandfold, Physical Chemistry Chemical Physics, 17 (2015) 24978-24987.
[52] X. Fu, D.A. Jacobs, F.J. Beck, H. Shen, K.R. Catchpole, T.P. White, Photoluminescence study of time- and spatial-dependent light induced trap de-activation in CH3NH3PbI3 perovskite films, Physical Chemistry Chemical Physics, 18 (2016) 22557-22564.
[53] W. Zhang, V.M. Burlakov, D.J. Graham, T. Leijtens, A. Osherov, V. Bulović, H.J. Snaith, D.S. Ginger, S.D. Stranks, Photo-induced halide redistribution in organic–inorganic perovskite films, Nature Communications, 7 (2016) 11683.
[54] Q. Chen, Y. Zhang, The reversal of the laser-beam-induced-current contrast with varying illumination density in a Cu2ZnSnSe4 thin-film solar cell, Applied Physics Letters, 103 (2013) 242104.
[55] A.M.A. Leguy, Y. Hu, M. Campoy-Quiles, M.I. Alonso, O.J. Weber, P. Azarhoosh, M. van Schilfgaarde, M.T. Weller, T. Bein, J. Nelson, P. Docampo, P.R.F. Barnes, Reversible Hydration of CH3NH3PbI3 in Films, Single Crystals, and Solar Cells, Chemistry of Materials, 27 (2015) 3397-3407.
[56] R. Treharne, A. Seymour-Pierce, K. Durose, K. Hutchings, S. Roncallo, D. Lane, Optical design and fabrication of fully sputtered CdTe/CdS solar cells, in:   Journal of Physics: Conference Series, IOP Publishing, 2011, pp. 012038.
[57] D.E. Aspnes, S.M. Kelso, R.A. Logan, R. Bhat, Optical properties of AlxGa1−x As, Journal of Applied Physics, 60 (1986) 754-767.
[58] J. Lee, N.C. Giles, D. Rajavel, C.J. Summers, Room-temperature band-edge photoluminescence from cadmium telluride, Physical Review B, 49 (1994) 1668-1676.
[59] Y. Zhang, B. Fluegel, M. Hanna, A. Duda, A. Mascarenhas, Electronic structure near the band gap of heavily nitrogen doped GaAs and GaP, Mat. Res. Soc. Symp. Proc., 692 (2002) 49.

[60] H.B. Bebb, E.W. Williams, Photoluminescence I: Theory in: R.K. Willardson, A.C. Beer (Eds.) Transport and Optical Phenomena, Academic Press, New York, 1972, pp. 181.
[61] N.T. Pelekanos, H. Haas, N. Magnea, H. Mariette, A. Wasiela, Room‐temperature exciton absorption engineering in II‐VI quantum wells, Applied Physics Letters, 61 (1992) 3154-3156.
[62] C. Vázquez‐López, H. Navarro, R. Aceves, M.C. Vargas, C.A. Menezes, Electroreflectance, photoreflectance, and photoabsorption properties of polycrystalline CdTe thin films prepared by the gradient recrystallization and growth technique, Journal of Applied Physics, 58 (1985) 2066-2069.
[63] M.A. Green, Y. Jiang, A.M. Soufiani, A. Ho-Baillie, Optical Properties of Photovoltaic Organic–Inorganic Lead Halide Perovskites, The Journal of Physical Chemistry Letters, 6 (2015) 4774-4785.
[64] Y. Zhang, B. Fluegel, M.C. Hanna, J.F. Geisz, L.W. Wang, A. Mascarenhas, Effects of heavy nitrogen doping in III–V semiconductors – How well does the conventional wisdom hold for the dilute nitrogen "III–V-N alloys"?, physica status solidi (b), 240 (2003) 396-403.
[65] M.E. Ziffer, J.C. Mohammed, D.S. Ginger, Electroabsorption Spectroscopy Measurements of the Exciton Binding Energy, Electron–Hole Reduced Effective Mass, and Band Gap in the Perovskite $CH_3NH_3PbI_3$, ACS Photonics, 3 (2016) 1060-1068.
[66] M. Shirayama, H. Kadowaki, T. Miyadera, T. Sugita, M. Tamakoshi, M. Kato, T. Fujiseki, D. Murata, S. Hara, T.N. Murakami, S. Fujimoto, M. Chikamatsu, H. Fujiwara, Optical Transitions in Hybrid Perovskite Solar Cells: Ellipsometry, Density Functional Theory, and Quantum Efficiency Analyses for $CH_3NH_3PbI_3$, Physical Review Applied, 5 (2016) 014012.
[67] A. Mascarenhas, Y. Zhang, The Physics of Tunable Disorder in Semiconductor Alloys, in: A. Mascarenhas (Ed.) Spontaneous Ordering in Semiconductor Alloys, Kluwer Academic/Plenum Publishers, New York, 2002, pp. 283.
[68] M.D. Sturge, Optical Absorption of Gallium Arsenide between 0.6 and 2.75 eV, Physical Review, 127 (1962) 768-773.
[69] M. Sebastian, J.A. Peters, C.C. Stoumpos, J. Im, S.S. Kostina, Z. Liu, M.G. Kanatzidis, A.J. Freeman, B.W. Wessels, Excitonic emissions and above-band-gap luminescence in the single-crystal perovskite semiconductors $CsPbBr_3$ and $CsPbCl_3$, Physical Review B, 92 (2015) 235210.
[70] F.O. Saouma, C.C. Stoumpos, M.G. Kanatzidis, Y.S. Kim, J.I. Jang, Multiphoton Absorption Order of $CsPbBr_3$ As Determined by Wavelength-Dependent Nonlinear Optical Spectroscopy, The Journal of Physical Chemistry Letters, 8 (2017) 4912-4917.
[71] P.Y. Yu, M. Cardona, Fundamentals of semiconductors: physics and materials properties, Springer, 2010.
[72] K. Miyata, T.L. Atallah, X.Y. Zhu, Lead halide perovskites: Crystal-liquid duality, phonon glass electron crystals, and large polaron formation, Science Advances, 3 (2017) e1701469.
[73] S. Sugai, Photoinduced tellurium precipitation in CdTe, Japanese Journal of Applied Physics, 30 (1991) L1083.
[74] T. Schmidt, K. Lischka, W. Zulehner, Excitation-power dependence of the near-band-edge photoluminescence of semiconductors, Physical Review B, 45 (1992) 8989-8994.
[75] X. Wen, Y. Feng, S. Huang, F. Huang, Y.-B. Cheng, M. Green, A. Ho-Baillie, Defect trapping states and charge carrier recombination in organic-inorganic halide perovskites, Journal of Materials Chemistry C, 4 (2016) 793-800.

[76] S. Draguta, S. Thakur, Y.V. Morozov, Y. Wang, J.S. Manser, P.V. Kamat, M. Kuno, Spatially Non-uniform Trap State Densities in Solution-Processed Hybrid Perovskite Thin Films, The Journal of Physical Chemistry Letters, 7 (2016) 715-721.
[77] M.I. Dar, G. Jacopin, S. Meloni, A. Mattoni, N. Arora, A. Boziki, S.M. Zakeeruddin, U. Rothlisberger, M. Grätzel, Origin of unusual bandgap shift and dual emission in organic-inorganic lead halide perovskites, Science Advances, 2 (2016).
[78] J.-C. Blancon, W. Nie, A.J. Neukirch, G. Gupta, S. Tretiak, L. Cognet, A.D. Mohite, J.J. Crochet, The Effects of Electronic Impurities and Electron–Hole Recombination Dynamics on Large-Grain Organic–Inorganic Perovskite Photovoltaic Efficiencies, Advanced Functional Materials, 26 (2016) 4283-4292.
[79] N.M. Haegel, T.J. Mills, M. Talmadge, C. Scandrett, C.L. Frenzen, H. Yoon, C.M. Fetzer, R.R. King, Direct imaging of anisotropic minority-carrier diffusion in ordered GaInP, Journal of Applied Physics, 105 (2009) 023711-023715.
[80] G.D. Gilliland, D.J. Wolford, T.F. Kuech, J.A. Bradley, Long-range, minority-carrier transport in high quality "surface-free" GaAs/AlGaAs double heterostructures, Applied Physics Letters, 59 (1991) 216-218.
[81] W. Li, S.K. Yadavalli, D. Lizarazo-Ferro, M. Chen, Y. Zhou, N.P. Padture, R. Zia, Subgrain Special Boundaries in Halide Perovskite Thin Films Restrict Carrier Diffusion, ACS Energy Letters, 3 (2018) 2669-2670.
[82] Y. Lin, Y. Zhang, Z. Liu, L. Su, J. Zhang, T. Wei, Z. Chen, Spatially resolved study of quantum efficiency droop in InGaN light-emitting diodes, Applied Physics Letters, 101 (2012) 252103.

Supporting information for

# Comparative studies of optoelectrical properties of prominent PV materials: Halide Perovskite, CdTe, and GaAs

*Fan Zhang[1], Jose F. Castaneda[1], Shangshang Chen[2], Wuqiang Wu[2], Michael J. DiNezza[3], Maxwell Lassise[3], Wanyi Nie[4], Aditya Mohite[5], Yucheng Liu[6], Shengzhong Liu[6], Daniel Friedman[7], Henan Liu[1], Qiong Chen[1], Yong-Hang Zhang[3], Jinsong Huang[2], and Yong Zhang[1*]*

[1]Department of Electrical and Computer Engineering, The University of North Carolina at Charlotte, Charlotte, North Carolina 28223, USA

[2]Department of Applied Physical Sciences, The University of North Carolina at Chapel Hill, Chapel Hill, North Carolina 27599, USA

[3]School of Electrical, Computer and Energy Engineering, Arizona State University, Tempe, Arizona 85287, USA

[4]Materials Physics and Application Division, Los Alamos National Laboratory, Los Alamos, NM 87545, USA

[5]Department of Chemical and Biomolecular Engineering and Department of Material Science and Nanoengineering Rice University, Houston, Texas 77005, USA

[6]Key Laboratory of Applied Surface and Colloid Chemistry, National Ministry of Education; Institute for Advanced Energy Materials, School of Materials Science and Engineering, Shaanxi Normal University, Xi'an 710062, China

[7]National Renewable Energy Laboratory, Golden, Colorado 80401, USA

*Correspondence: yong.zhang@uncc.edu

1. **Effect of numerical aperture in confocal PL measurement**

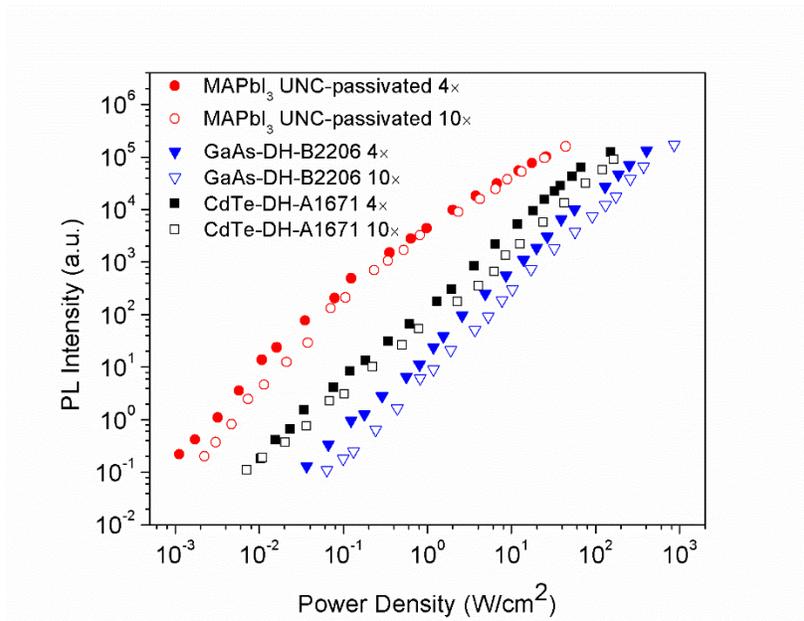

FIG.S1. PL intensity vs. excitation density using two microscope lenses: 4× with NA = 0.1, and 10× with NA = 0.25. Results are shown for three samples: MAPbI$_3$ (UNC-passivated), CdTe (CdTe-DH-A1671), and GaAs (GaAs-DH-B2206).

2. **PL intensity time dependence**

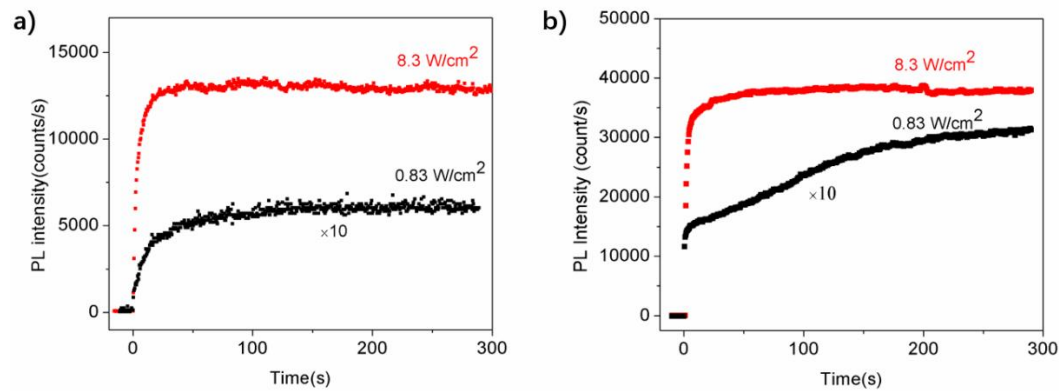

FIG.S2. PL intensity time map of thin film perovskite under 10× lens. (a) MAPbI$_3$-LANL at 0.83 W/cm$^2$ and 8.3 W/cm$^2$; (b) MAPbI$_3$-UNC-passivated at 0.83 W/cm$^2$ and 8.3 W/cm$^2$.

## 3. Absorption spectra of thin-film perovskite MAPbI₃

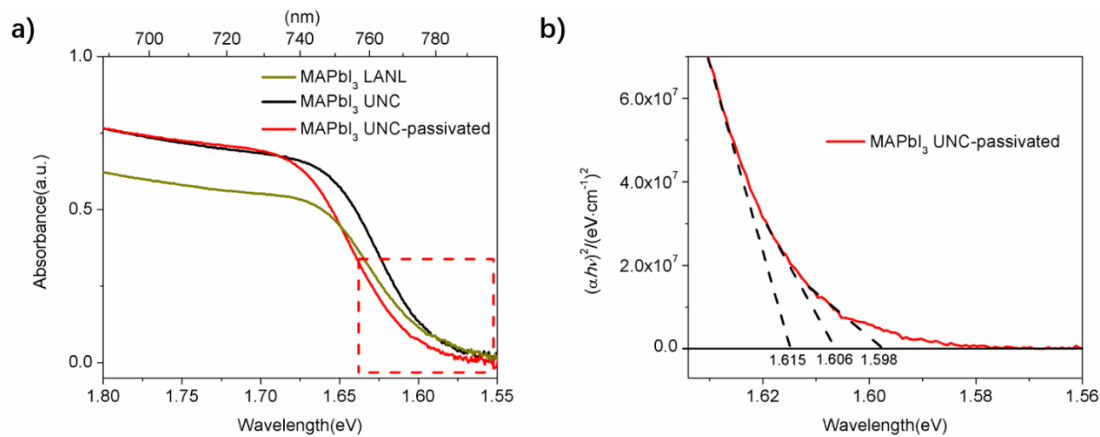

FIG.S3. (a) Absorption spectra of thin film perovskite measured by 10× lens. Reflection has been considered. (b) Tauc plot of UNC-passivated sample, corresponding to red dashed box in Figure S3(a). Dashed lines show possible extrapolated bandgaps.

## 4. Photon-recycling effect

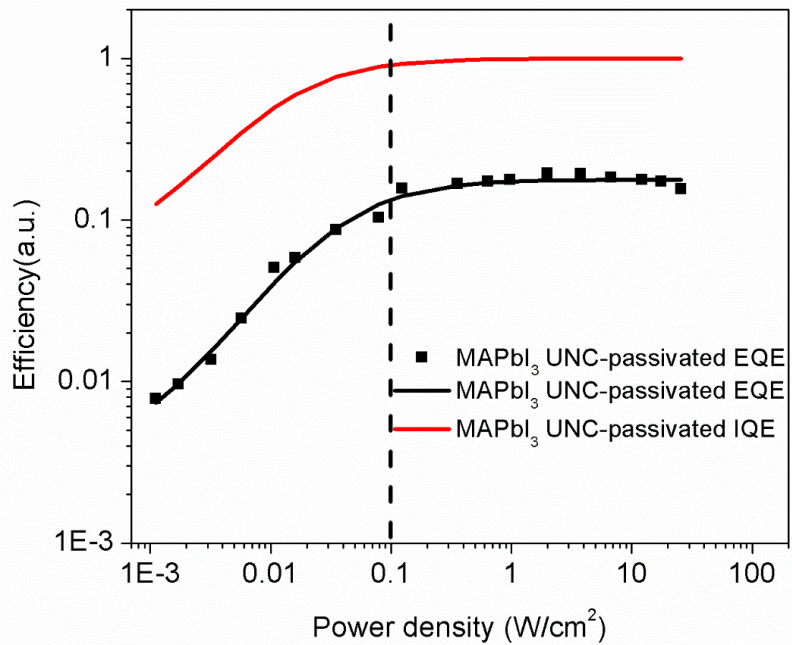

FIG.S4. Excitation density dependent external quantum efficiency for MAPbI₃ sample UNC-passivated. Black solid lines are fitted curve and vertical dashed line shows one sun illumination. Red line is the internal quantum efficiency for the same data.

We have pointed out in the main text that the photo-recycling effect unlikely plays a significant role in the measured EQE. However, we do wish to examine what if the photo-recycling were important the impact would be to the extracted IQE. We use "UNC-passivated" as an example to compare the results of the two assumptions.

Internal quantum efficiency $\eta_{IQE}$ can be written as below:

$$\eta_{IQE} = \frac{1}{2}\left(1 - \beta_0'(1+dx^\zeta)\frac{u(1+dP^\zeta)+1}{P} + \sqrt{(1+\beta_0'(1+dP^\zeta)\frac{u(1+dP^\zeta)+1}{P})^2 - \frac{4\beta_0'(1+dP^\zeta)}{P}}\right). \quad (S1)$$

Eq. (S1) is the same formula as Eq. (5) in main text for relative external quantum efficiency $\eta_{EQE}$, except for removing the constant C. Now measured relative external efficiency $\eta_{EQE}$ can be linked to $\eta_{IQE}$ in the following way:

$$\eta_{EQE} = \frac{\frac{C}{2n_r^2}\eta_{IQE}}{\frac{1}{2n_r^2}\eta_{IQE} + 1 - \eta_{IQE} + \frac{L}{4\alpha_0 d_0}} \quad . \quad (S2)$$

Eq. (S2) adopts the relationship between IQE and absolute EQE proposed in Ref. [1], where C is a scaling factor because we do not measure the absolute EQE. $n_r$ is the average refractive index at the perovskite PL peak position, $\alpha_0$ is the average band-edge absorption coefficient over the perovskite emission band, $d_0$ is the absorber thickness, L is the loss factor, defined as L = 1 − Reflectivity. Eq. (S2) can be used to fit the experimental data of relative EQE. The absolute EQE can then be calculated by dividing fitting curve of Eq. (S2) and the experimental data with C. Taking the same value of $n_r$ = 2.65, and the same $\alpha_0$ value as in Ref. [2], but adjusted for the thickness difference (in our case $d_0$ = 500 nm), yielding $\alpha_0 d_0$ = 0.6, and L = 0.796. The fitting results of the IQE and absolute EQE curve are shown in Fig. S4, which yields a EQE of 60% (normalized) and IQE of 85%, compared to the reported 35% EQE (normalized) and 92% IQE for a TOPO capped sample, under 1 Sun equivalent (60 mW/cm$^2$ at 532 nm) [2]. In the main text where photo-recycling is not considered, we obtain an IQE of 60% under 1 Sun equivalent.

[1] I. Schnitzer, E. Yablonovitch, C. Caneau, T.J. Gmitter, Ultrahigh spontaneous emission quantum efficiency, 99.7% internally and 72% externally, from AlGaAs/GaAs/AlGaAs double heterostructures, Applied Physics Letters, 62 (1993) 131-133.
[2] I.L. Braly, D.W. deQuilettes, L.M. Pazos-Outón, S. Burke, M.E. Ziffer, D.S. Ginger, H.W. Hillhouse, Hybrid perovskite films approaching the radiative limit with over 90% photoluminescence quantum efficiency, Nature Photonics, 12 (2018) 355-361.